\documentclass[
aps,%
12pt,%
final,%
notitlepage,%
oneside,%
onecolumn,%
nobibnotes,%
nofootinbib,%
superscriptaddress,%
noshowpacs,%
centertags]%
{revtex4}

\newcommand{\beq}{\begin{equation}}
\newcommand{\eeq}{\end{equation}}
\newcommand{\bea}{\begin{eqnarray}}
\newcommand{\eea}{\end{eqnarray}}

\def\BB{{\cal B}}
\providecommand{\dif}{\mathrm{d}} \def\d{\dif}

\begin{document}
\selectlanguage{english}

\title{Constraints on mass, spin and magnetic field of microquasar H~1743-322 from observations of QPOs}

\author{\firstname{A. A.}~\surname{Tursunov}}
\email{arman.tursunov@fpf.slu.cz}
\affiliation{
Institute of Physics, Faculty of Philosophy and Science, Silesian University in Opava, \\
Bezru{\v c}ovo n{\'a}m.13, CZ-74601 Opava, Czech Republic
}
\affiliation{Bogoliubov Laboratory of Theoretical Physics,\\
Joint Institute for Nuclear Research, 141980 Dubna, Russia
}
\author{\firstname{M.}~\surname{Kolo\v{s}}}
\email{martin.kolos@fpf.slu.cz}
\affiliation{
Institute of Physics, Faculty of Philosophy and Science, Silesian University in Opava, \\
Bezru{\v c}ovo n{\'a}m.13, CZ-74601 Opava, Czech Republic
}

\begin{abstract}
The study of quasi-periodic oscillations (QPOs) of X-ray flux observed in many microquasars can provide a powerful tool for testing of the phenomena occurring in strong gravity regime. QPOs phenomena can be well related to the oscillations of charged particles in accretion disks orbiting Kerr black holes immersed in external large-scale magnetic fields. In the present paper we study the model of magnetic relativistic precession and provide estimations of the mass and spin of the central object of the microquasar H~1743-322 which is a candidate for a black hole. Moreover, we discuss the possible values of external magnetic field and study its influence on the motion of charged particles around rotating black hole.
\end{abstract}

\maketitle

\section{Introduction}

Microquasar is binary system composed of a compact object, such as a black hole, and a companion (donor) star. Matter floats from the companion star onto the black hole trough some rotating axially symmetric structure called accretion disk. The friction of a matter of an accretion disk leads to the emission of electromagnetic radiation in the wide range of spectra including X-rays. One of the promising tools to understand the phenomena occurring in the strong gravity regime is the study of the quasi-periodic oscillations (QPOs) of the X-ray power density observed in black hole microquasars containing black hole candidates. The current technical possibilities to measure the frequencies of QPOs with high precision allow us to get some useful knowledge about the parameters of black holes and their vicinities. Depending on the frequencies, which cover the range from few mHz up to $0.5$kHz, different types of QPOs were distinguished. Mainly, these are the high frequency (HF) and low frequency (LF) QPOs with frequencies up to 500 Hz and up to 30 Hz, respectively. The HF QPOs in black hole microquasars are sometimes detected with the twin peaks, which have a frequency ratio close to $3:2$.

Test particle revolving around black hole can undergo quasi-harmonic oscillations around stable circular orbits. The frequencies of such oscillations can be comparable with the observed frequencies of QPOs.
 Since, the matter surrounding black hole contain also charged particles, the influence of external magnetic fields on their dynamics cannot be neglected. 
Thus, the minimal assumptions of proposed model are that the observed QPOs are produced by charged particles in the vicinity of a black hole immersed into external large-scale magnetic field, which we suppose to be asymptotically uniform. In the present paper we examine the influence of such fields on the QPOs phenomena determining mass and spin of the black hole candidate of the H~1743-322 source. Charged particle motion in a uniform magnetic field and related high frequency oscillations have been studied for Schwarzschild black holes in \cite{Kol-Stu-Tur:2015:CLAQG:}. Locally measured angular frequencies for the magnetized Kerr black hole with the dynamical equations for charged particles were found in \cite{Tur-Stu-Kol:2016:PRD:}. Estimations of magnetic fields in the vicinity of three other microquasars GRS 1915+105, XTE 1550-564, and GRO 1655-40  have been performed in \cite{Kol-Tur-Stu:2017:submitted}. Hereafter we mainly use the methods developed in above mentioned papers for the construction of parameters of the black hole candidate H~1743-322.

\section{Charged particle dynamics}

There are many evidences that magnetic fields have to be present in the galaxies. For example, the equipartition magnetic field strength in the center of the Galaxy is estimated to be around $\sim 10 {\rm{G}}$ \citep{Eat-etal:2013:NATUR:}. A strength of magnetic field decreases with the distance from the center achieving $\sim 10^{-5}{\rm{G}}$ at the edge.

The relative strength of the Lorentz force acting on a charged particle moving around weakly magnetized black hole is represented by the following dimensionless parameter
\beq
 \quad \BB = \frac{q B G M}{2 m c^4}, \label{BBdef}
\eeq
where $q$ and $m$ are the charge and mass of a test particle; $M$ is a mass of a black hole, $G$ and $c$ are the world constants in their usual notations. Further we call $\BB$ as the magnetic parameter which characterizes the relative strengths of magnetic and gravitational fields.
The estimation of the value of magnetic parameter $\BB$ in realistic situations shows that the effect of even weak magnetic field cannot be neglected for the motion of  charged particle due to the large value of the specific charge $q/m$ of a test particle.
Following to \cite{Kol-Tur-Stu:2017:submitted}, the oscillatory frequencies of charged particles in black hole surroundings filled with asymptotically uniform magnetic field can be well fitted with the frequencies of QPOs observed in the microquasars GRS 1915+105, XTE 1550-564 and GRO 1655-40. Corresponding values of the magnetic parameter in these cases are $|\BB|\sim 0.004$. However, the strength of magnetic field (in Gauss) corresponding to the magnetic parameter $|\BB|\sim 0.004$ can have completely different magnitudes depending on the type of charged particle, i.e. its specific charge $q/m$ and the black hole mass. For example, for stellar mass black hole $M={10 {M}_{\odot}}$, the parameter $|\BB|\sim 0.004$ corresponds to the magnetic field strength $B = 10^{-5}~{\rm{G}}$ for electron, $B = 1 {\rm{G}}$ for partially ionized Ferrum atom (Fe+) or $B = 10^{9}~{\rm{Gs}}$ for charged dust grain with one electron lost ($m=2\times10^{-16}$~kg). This implies that even for small magnetic parameter $|\BB|=0.004$, the Lorentz force is not negligible compared to the gravitational force of the central black hole \cite{Kol-Tur-Stu:2017:submitted}.

The motion of a charged particle with charge $q$ and mass $m$ in curved spacetime in the presence of electromagnetic field is described by the following equation
\beq
 \frac{\d u^\mu}{\d \tau} + \Gamma_{\alpha\beta}^{\mu} u^{\alpha} u^{\beta} = \frac{q}{m} g^{\mu\rho} F_{\rho\sigma} u^{\sigma}, \label{geomageq}
\eeq
where $u^{\mu} = dx^{\mu} / d\tau$ is the four-velocity of a particle, $\Gamma_{\alpha\beta}$ are Christoffel symbols for Kerr black hole metric and $F_{\mu\nu}$ is a tensor of electromagnetic field. The four velocity of the circular motion in equatorial plane has only 2 non-vanishing components, $u^{\mu} = \{u^{t},0,0,u^{\phi}\}$. The radial component of Eq. (\ref{geomageq}) can be written in the form
\beq
(a^2 - r^3) (u^{\phi})^2 - 2 \BB (r^3 - a^2) u^{\phi} -  2 a (u^{\phi} + \BB) u^{t} + (u^{t})^2 = 0. \label{geomageqR}
\eeq
The normalization condition $g_{\mu\nu} u^{\mu} u^{\nu} = -1$ gives the second equation for the nonzero components of the four-velocity in the form
\beq
(a^2 (2 + r) + r^3) (u^{\phi})^2 - 4 a u^{\phi} u^{t} - (r-2) (u^{t})^2 + r = 0, \label{norcon}
\eeq
which is obviously independent of the electromagnetic parameter $\BB$. Equations (\ref{geomageqR}) and (\ref{norcon}) allow us to find two expressions for two unknown quantities $u^{t}$ and $u^{\phi}$. 
The analytical form of expressions for $u^{t}$ and $u^{\phi}$ cannot be represented in a reasonable form, however can be easily solved numerically.

Following \cite{Ali-Gal:1981:GRG:,Kol-Tur-Stu:2017:submitted}, we consider the oscillatory motion of charged particle and use the method of perturbation of the equations of motion around the stable circular orbits. Then the explicit form of equations for the radial and vertical oscillatory frequencies take the following form
\bea
 \Omega_\theta^2 &=& \frac{\alpha u^\phi (u^\phi + 2 \BB) + \beta u^t (u^\phi + \BB) + 2 a^2 (u^t)^2}{r^5 (u^t)^2} , \label{Omtheta} \\
 \Omega_r^2 &=& \frac{\gamma u^\phi (u^\phi + 2 \BB) + \mu u^t (u^\phi + \BB) + \rho (u^t)^2 + \sigma}{r^5 (u^t)^2}, \label{Omradial}
\eea
where the introduced coefficients are
\bea
\alpha &=& r^5 + a^2 r^2 (4 + r) + 2 a^4, \nonumber \\
\beta &=& -4 a (r^2 + a^2), \nonumber \\
\gamma &=& r^4 (-8 + 3 r) + a^2 r (2 - 10 r + r^2) - 4 a^4, \nonumber \\
\mu &=& 4 a (2 a^2 - r + 3 r^2), \nonumber \\
\rho &=& 2 (-2 a^2 + r - r^2), \nonumber \\
\sigma &=& 4 \BB^2 r ((-2 + r) r^3 - a^2 (1 + 2 r)).
\eea
Here $u^\phi$ and $u^{t}$ are the components of the four velocity of a particle given by the solution of Eq. (\ref{geomageqR}) and (\ref{norcon}). In addition to the epicyclic frequencies $\Omega_r$ and $\Omega_{\theta}$, there exist also the Keplerian frequency, $\Omega_{\rm K}$, and so called Larmor angular frequency, $\Omega_{L}$, associated with the pure contribution of an external uniform magnetic field and the gravitational redshift. The frequencies $\Omega_{\rm K}$ and $\Omega_{L}$ take the form
\beq
\Omega_{\rm K} \equiv \Omega_{\phi} = \frac{d \phi}{d t} = \frac{u^\phi}{u^{t}}, \qquad \Omega_{L} \equiv \frac{q B}{m u^t} = \frac{2\BB}{u^t}. \label{OmKepLar}
\eeq
The frequency $\Omega_{L}$ is fully relevant in large distances from the black hole where influence of the uniform magnetic field becomes crucial and epicyclic oscillations vanish.

The expressions for the fundamental frequencies (\ref{Omtheta}), (\ref{Omradial}) and (\ref{OmKepLar}) are given in the dimensionless form. In CGS units, one needs to extend the corresponding formulas by the factor $c^3/GM$. Then the radial and latitudinal frequencies of the harmonic oscillations of a charged particle measured by observer at rest at infinity in Hz are given by
\beq
     \nu_{i} = \frac{1}{2\pi} \frac{c^3}{GM} \, \Omega_{i} [{\rm Hz}], \label{nu_rel}
\eeq
where $i\in\{r,\theta,\phi\}$. Detailed analysis of the frequencies $\nu_{i}$ can be found in \cite{Kol-Tur-Stu:2017:submitted}.

\section{Model of magnetic relativistic precession} \label{observations}

The HF QPOs come in pairs of two peaks with upper $f_{\mathrm{U}}$ and lower $f_{\mathrm{L}}$ frequencies in the timing spectra, with the frequency ratio $f_{\mathrm{U}}:f_{\mathrm{L}}$ very close to the fraction of $3:2$. Observations of this effect in different non-linear systems show the existence of the resonances between two modes of oscillations.
Simultaneous detections of twin HF QPOs and single LF QPOs enables us to obtain the stringent restrictions on the mass and dimensionless spin of the central black hole, if we assume that all of these QPOs arise at a given radius of the accretion disk \citep{Mot-etal:2014a:MNRAS:}. In this sense the advantages of black holes are that they are described by only two parameters -- mass and spin.

In case of geodesic QPO models, the observed frequencies are associated with the different linear combinations of the particle fundamental frequencies $\nu_{r}, \nu_{\theta}$ and $\nu_{\phi}$. One of such models is the model of relativistic precession (RP) introduced in \citep{Ste-Vie-Mor:1999:ApJ:}. The RP model is usually applied for the explanation of simultaneously detected HF and LF QPOs observed in X-ray binaries containing black hole or neutron stars. Despite the fact that the original RP model which assumes the oscillations of neutral matter have demonstrated its practical usefulness, it is not able to fit the observational QPO data for all microquasars simultaneously. However in \cite{Kol-Tur-Stu:2017:submitted} it was pointed out that the electromagnetic interaction can play important role and the tripple QPOs (two HF and one LF) for three Galactic microquasars were successfully fitted within so called magnetic relativistic precession model (MRP).
In the MRP model the double HF ($\nu_{\rm U}$ and $\nu_{\rm L}$) and single LF ($\nu_{\rm low}$) QPOs are identified with the following frequencies of a test particle
\beq
 \nu_{\rm U}=\nu_{\phi}, \quad \nu_{\rm L}=\nu_{\phi}-\nu_{\rm r}, \qquad \nu_{\rm low}=\nu_{\phi}-\nu_{\theta}, \label{RPmodel2}
\eeq
where $\nu_{\phi}$ is the orbital frequency of a particle, $\nu_{\phi}-\nu_{\rm r} \equiv \nu_{per}$ is the frequency of the perihelion precession, and $\nu_{\phi}-\nu_{\theta} \equiv \nu_{nod}$ is the nodal precession frequency related to the non-spherical symmetry of a rotating black hole.

While the upper frequency of HF QPOs for the microquasar H~1743-322 has been measured quite precisely, the lower frequency of HF QPOs has large discrepancy in the  measurements presented in a literature. The largest discrepancy in lower frequency $f_{\rm L}$ to the date corresponds to the following values 
 given in Hz
\bea
  && f_{\rm U} = 240\pm3, \quad f_{\rm L} = 165\pm5, \quad f_{\rm L} = 9.44\pm0.02 \label{data1} \\
  && f_{\rm U} = 242\pm3, \quad f_{\rm L} = 166\pm5, \quad f_{\rm L} = 22 \label{data2}
\eea
see Tab. 1. in \cite{Ing-Mot:2014:MNRAS:} and Tab. 2. in \cite{Ste:2014:MNRAS:}. One can use the simultaneously observed values of HF and LF QPOs $f_{\rm U}, f_{\rm L}, f_{\rm low}$ and identify them with formula for MRP model (\ref{RPmodel2}) as three equations
\beq
  f_{\rm U}=\nu_{\rm U}(M,a,\BB), \quad f_{\rm L}=\nu_{\rm L}(M,a,\BB), \quad f_{\rm low}=\nu_{\rm low}(M,a,\BB) \label{RPmodel3}
\eeq
for three independent parameters - black hole mass $M$, spin $a$ and magnetic parameter $\BB$. 

In addition to three static parameters $M$, $a$ and $\BB$, the frequencies in MRP model depend also on the location of the particle orbit $r$ where the resonances occur. Each set of static parameters for observed QPO frequencies allow us to find the location of the QPO resonances $r$ in the black hole vicinity. Since for H~1743-322 source, none of the parameters are precisely determined (from any kind of independent measurements) one can restrict the parameter $\BB$ on its possible range $\BB\in(-0.004,0.004)$, which as it was stated above, covers the range of astrophysically relevant values of magnetic fields for Galactic microquasars. This allows us to find the range of possible values of the black hole mass $M$ and spin $a$. The region of allowed black hole mass $M$ and spin $a$ for H~1743-322 source has been obtained by fitting the observed frequencies as demonstrated in the figure.

\section{Results and conclusions} \label{kecy}

We have studied the influence of external weak magnetic field on the motion of charged particle in its relation to the explanation of QPOs recently detected in H~1743-322 source. Predicted mass and dimensionless spin for central black hole using magnetic field parameter $|\BB|<0.004$ are $M = (7.4-11.2) M_{\odot}$, $a/M = (0.2-0.4)$ for frequencies (\ref{data1}) and $M = (11.2 - 13.6) M_{\odot}$, $a/M = (0.4-0.6)$ for frequencies (\ref{data2}). Note, that the restriction of the magnetic parameter in the range $|\BB|<0.004$ corresponds to the astrophysically relevant values of the magnetic fields for elementary particles.

It is interesting to note, that both set of observed frequencies (\ref{data1}) and (\ref{data2}) can be explained within single model of charged particles orbiting magnetized Kerr black hole if to suppose that two independent observations correspond to QPOs from charges with different signs: $q>0$ for (\ref{data1}) and $q<0$ for (\ref{data2}). In that case, the mass and the spin of the central object of H~1743-322 correspond to the values $M = 11.2 M_{\odot}$ and $a = 0.4$. In other words, both QPO measurements could be correct in case if they are generated by the particles with different signs.

Remarkably, the mass of H~1743-322 source has been estimated recently in \cite{Mol-etal:2017:APJ:} as $M = 11.2 M_{\odot}$ using different technique.

\begin{acknowledgments}

The authors acknowledge the Silesian University in Opava Grant No. SGS/14/2016 and the Czech Science Foundation Grant No. 16-03564Y.

A.T. acknowledges the kind hospitality of Bogoliubov Laboratory of Theoretical Physics of the Joint Institute for Nuclear Research in Dubna.

\end{acknowledgments}


\def\mnras{Mon. Not. R. Astron Soc.}

\selectlanguage{english}

\begin{figure}[t!]
\setcaptionmargin{5mm}
\onelinecaptionstrue  
\includegraphics[width=\hsize]{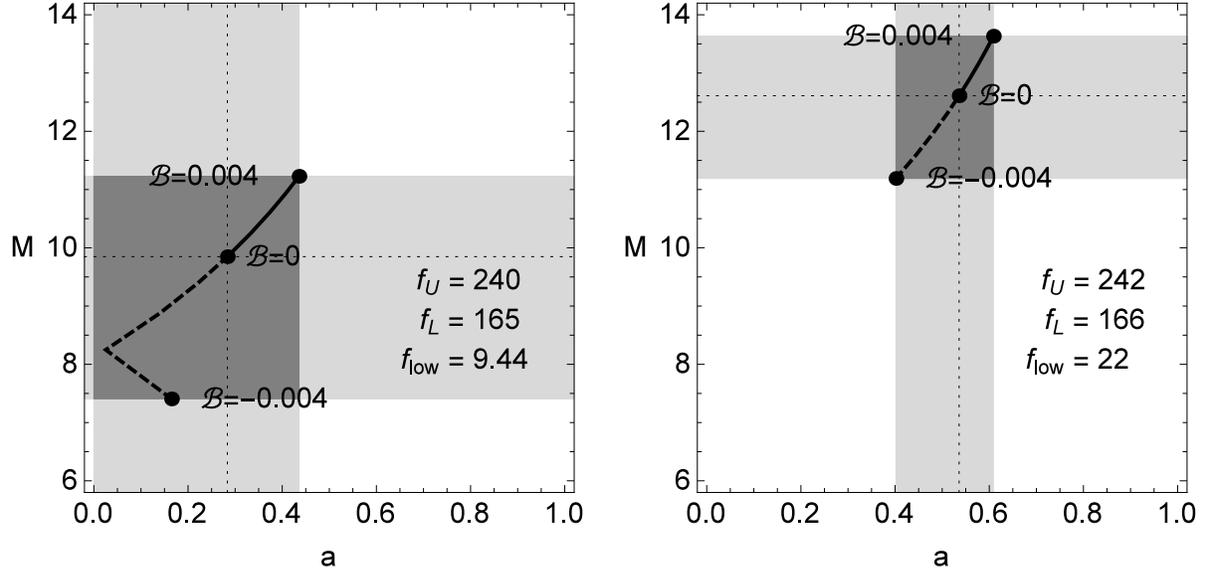}
\captionstyle{normal}
\caption{\label{fig1}
Constraints of black hole mass $M$ (in the units of solar mass) and dimensionless spin $a$ for microquasar H~1743-322 given by MRP model for two sets of observed frequencies with largest discrepancy. Observed frequencies for HF ($f_{\rm U}$ and $f_{\rm L}$) and LF ($f_{\rm low}$) QPOs used for the fitting are shown inside each figure.
}
\end{figure}

\end{document}